\documentstyle[12pt,epsf]{article}
\newcommand{\Zset}{{\sf Z}\hskip -5.5pt {\sf Z}}

\newcommand{\cA}{{\cal A}}

\newcommand{\bi}{\bigskip}
\newcommand{\no}{\noindent}
\newcommand{\hk}{\hspace{0.1cm}}
\newcommand{\hs}{\hspace{0.5cm}}
\newcommand{\dslash}{\partial\hskip -0.5em/}

\newcommand{\rk}{\right)}
\newcommand{\lk}{\left(}

\newcommand{\il}{\int\limits}

\newcommand{\be}{\begin{equation}} 
\newcommand{\en}{\end{equation}}
\newcommand{\bea}{\begin{eqnarray}}
\newcommand{\ena}{\end{eqnarray}}

\newcommand{\hbo}{\hbox to 1 true cm {\hfill } } 
\newcommand{\tr}{\hbox{tr}}

\def\dslash{\partial\kern-.5em\slash}
\def\kslash{k\kern-.5em\slash}
\def\pslash{p\kern-.5em\slash}
\def\rr{\hbox{R}\kern-1.2em\mid \; }

\begin{document} 
\vglue 1truecm
  
\vbox{ UNITU-THEP-20/1996 
\hfill October 25, 1996
}
  
\vfil
\centerline{\large\bf Monopoles and Strings in Yang-Mills Theories$^*$ } 
  
\bigskip
\centerline{ K.~Langfeld, H.~Reinhardt, M.~Quandt$^1$ } 
\bigskip
\vspace{1 true cm} 
\centerline{ Institut f\"ur Theoretische Physik, Universit\"at 
   T\"ubingen }
\centerline{D--72076 T\"ubingen, Germany.}
\bigskip
\vskip 1.5cm

\begin{abstract}
\noindent 
Yang-Mills theory is studied in a variant of 't Hooft's maximal
Abelian gauge. In this gauge magnetic monopoles arise in the
Abelian magnetic field. 
We show, however, that the full (non-Abelian) magnetic field
does not possess any monopoles, but rather strings of magnetic
fluxes. We argue that these strings are the relevant infrared degrees 
of freedom. The properties of the magnetic strings which arise 
from a dilute instanton gas are investigated for the gauge group SU(2).

\end{abstract}

\vfil
\hrule width 5truecm
\vskip .2truecm
\begin{quote} 
$^*$ Supported in part by DFG under contract Re 856/1-3. 

$^1$ Supported by Graduiertenkolleg "Hadronen und Kerne" 

\end{quote}
\eject

\no
\bi

\section{Introduction } 
Confinement may be realized through a dual Meissner effect~\cite{tho76}. 
This confinement scenario assumes that the QCD ground state consists
in a certain gauge of a condensate of magnetic monopoles. Such a dual
super-conductor squeezes the color electric field between color
charges into flux tubes and in this way provides confinement. 
This confinement scenario seems to be realized in super-symmetric 
models~\cite{sei94}. 

Although this idea dates back almost twenty years, recent developments 
in numerical simulations of lattice gauge theory provide the environment 
to test such ideas in a quantitative manner. 
The main ingredient in a derivation of the dual Meissner effect 
is the Abelian projection. Let us briefly illustrate this procedure for 
later use. The starting point is the Cartan decomposition of the 
gauge group $G= H \times G/H $ where $H$ denotes the Cartan subgroup. 
Accordingly, the link variable $U_\mu (x) = u_\mu (x) M_{\mu }(x)$ 
is decomposed into an Abelian part $u_\mu (x) $ and a coset part 
$M_\mu (x)$. The latter contains the gluonic field $A^{ch}_\mu $, which 
is charged with respect to the Cartan-group. Lattice calculations provide 
support for the notion that in the so-called maximal Abelian 
gauge~\cite{kro87}, the Abelian field components and in particular the 
magnetic monopoles dominate the infrared physics (Abelian 
dominance)~\cite{pol95} and that those charged gluon fields 
$A^{ch}_\mu $ which are assumed to be free of topological 
obstructions, can be perturbatively taken into account. 
In particular, for the tension of the 
string connecting static sources in the fundamental representation, 
the Abelian lattice configurations reproduce about 
92\% of the string tension and furthermore 95\% of this part
comes from the magnetic monopoles alone~\cite{r5}. 

Despite the success at hand, recent studies cast doubt onto the 
widespread belief in Abelian dominance. From 
the analytical approach to finite temperature Yang-Mills theory 
of~\cite{dam95}, supplemented by numerical results, the authors 
concluded that the Abelian projection provides correct qualitative 
features, whereas it fails at a quantitative level. To be more 
precise, lattice measurements reveal that the tension of the string 
between static sources of representation $R$ is proportional 
to the quadratic Casimir $C_R$~\cite{fab88} (Casimir scaling). 
The Abelian dominance 
approximation obviously fails to reproduce this observation, since 
e.g. quarks in the adjoint representation possess 
Abelian charge zero, and hence the string tension in the Abelian 
approximation is zero in contradiction to the full lattice 
result~\cite{deb96}. This result indicates that the contributions of 
the charged 
fields $A^{ch}_{\mu }$ to physical quantities are not negligible  
even in the infra red. This fact is confirmed by a large-$N$ 
analysis of SU(N) Yang-Mills theory~\cite{deb94}. It was 
observed that the forces between Abelian--charged and between 
Abelian--neutral sources possess the same order of magnitude. 
In this paper, we will provide further evidence that the charged 
gauge fields $A^{ch}_\mu$ contribute to the partition function 
in a topological by non-trivial manner. 

There is also evidence from other lattice calculations that the 
magnetic monopoles obtained in the maximal Abelian gauge need not 
be the most efficient infra red degrees of freedom. In the lattice study 
reported in~\cite{deb96}, the Abelian gauge field is brought as close 
as possible to a $Z_2$ gauge field by fixing the residual U(1) 
gauge symmetry. Assuming that this gauge field is the only 
relevant infra red degree of freedom (center dominance), the 
string tension of quarks in the fundamental representation\footnote{The 
center dominance approximation also fails to reproduce Casimir 
scaling~\cite{deb96}. } was reproduced to high accuracy. 
This result suggests that the QCD vacuum is as well described in terms 
of vortices as by a condensate of monopoles. 

In this paper we will show in a variant of 't Hooft's  maximal 
Abelian gauge that the emergence of the magnetic monopoles is  
confined to the Abelian magnetic field $\vec{\cal B}^a = \hbox{rot} 
\vec{A}^a $, and therefore highly 
gauge dependent. We will find that the full (non-Abelian) magnetic 
field strength does not contain any monopoles. The Abelian magnetic 
monopole field is canceled by the commutator contribution 
$[A^{ch}_\mu , A^{ch}_\nu ]$ to the field strength tensor. 
The full field strength, which is the relevant 
quantity in the partition function, possesses 
magnetic strings, which are the leftovers of the Dirac strings of
the Abelian monopole field. We further argue that it is these
strings rather than the magnetic monopoles which are the dominant
infrared degrees of freedom. Finally, we discuss in detail the strings 
occurring in a dilute instanton medium.  

The paper is organized as follows: in the next section, we will 
discuss the emergence of magnetic monopoles as artifacts of incomplete 
gauge fixings which leave a residual Abelian invariance. 
For a particular choice of the Abelian gauge fixing, 
we extract the magnetic string attached to the monopole. 
In section 3, we will show that the monopoles drop out from the full 
field strength while the string survives. We will argue that these 
strings are the relevant infra-red degrees of freedom. In section 4, 
the distribution of the string length will be investigated in a dilute 
instanton gas. Our conclusions will be left to the final section.

\section{Topological Properties of monopoles }

\noindent
In what follows we briefly explain the emergence of magnetic monopoles 
in Abelian gauges and discuss their topological 
properties~\cite{tho76,kro87}. 
We will find that the charges of these monopoles are quantized on 
topological grounds. Furthermore we will clarify in which respect 
these monopoles differ from the magnetic monopoles in classical 
electro dynamics. 

Consider a field ${\bf \Phi } (x) = \Phi ^a (x) T^a \equiv 
\vec{\Phi }(x) \vec{T}$, 
which lives in the Lie algebra of the gauge group $G=SU(N)$ 
and transforms homogeneously under gauge transformations
\be
\label{E1}
{\bf \Phi } \longrightarrow {\bf \Phi }^\Omega = \Omega {\bf \Phi } \Omega ^\dagger \hk . 
\en
We can fix the gauge by diagonalizing this field
\be
\label{E2}
{\bf \Phi }^\Omega = \Lambda (x) = \Lambda^{C_0} T^{C_0} \hs , \hs {\bf \Phi } =
\Omega^{- 1} \Lambda \Omega \hk , 
\en
where $\Lambda (x) = diag \lk \lambda_1, \dots \lambda_N \rk$ is a
diagonal matrix with $\lambda_i > \lambda_{i + 1}$ 
living in the Cartan sub-algebra. Mathematically, eq.
(\ref{E2}) means that the field ${\bf \Phi }$ is conjugated into the maximal
torus. Eq. (\ref{E2}) does not fix the gauge uniquely but still leaves 
invariance under transformations of the Cartan subgroup 
$H=[U(1)]^{N-1}$. Consequently the
gauge function $\Omega (x)$ is defined only up to an element of the
Cartan subgroup. In fact the transformation $\Omega (x) \to t (x) \Omega
(x)$ with $t (x) \in H$ does not change the field ${\bf \Phi }$. Consequently
the gauge transformation $\Omega (x)$ can be restricted to the coset
space $G / H$. At those points $x = \bar{x}$ where two eigenvalues
of ${\bf \Phi } (x)$ coincide, the gauge function $ \Omega (x)$ is not well
defined. Without loss of generality we can arrange the eigenvalues of
${\bf \Phi }$ in such a way that the two degenerate eigenvalues correspond to
a $SU (2)$ subgroup. In this case, it is obvious that these eigenvalues 
have to vanish at the degeneracy point $\bar{x}$ . For simplicity let us 
confine ourselves in the following to the $SU (2)$ gauge group. 
Throughout this paper, we use anti-hermitian generators $T^{a}= -i \tau ^a 
/2 $, where $\tau ^a$ are the Pauli matrices. 
The gauge transformation $\Omega $ which diagonalizes ${\bf \Phi } $ can be 
chosen as 
\be 
\Omega \; = \; - i \hat{\alpha } \vec{\tau } \; , \hbo 
\hat{\alpha }^2 =1 \; , 
\label{eq:4} 
\en 
where the unit color vector $\hat{\alpha }$ is given by 
\be 
\hat{ \alpha }^{a} \; = \; \frac{ \Phi \delta ^{a3} + \Phi ^a }{ 
\sqrt{ 2 \Phi ( \Phi ^3 +\Phi ) }} \; , \hbo 
\Phi := \sqrt{ \vec{\Phi }^2 } \; . 
\label{eq:5} 
\en 
Then the 
degeneracy points correspond to vanishing field configurations $\Phi ( x
= \bar{x} ) = 0$. Near these degeneracy points the field hence has the
form
\be
\label{E3}
\Phi^a = C^{a}_{i} (x - \bar{x} )_i  
\en
with some constants $C^a_i$. By a coordinate transformation 
$x^{' \, a } = C^a_i x^i$, this configuration can be brought to the 
hedgehog form $\Phi ^a = (x'-\bar{x}')^a$. As shown by 
't~Hooft~\cite{tho76} (see also~\cite{kro87,sug95}), the gauge 
transformation which diagonalizes the hedgehog field is such that the 
transformed gauge potential 
\be
\label{E4}
A^{\Omega}_{i} = \Omega A_i \Omega^\dagger + \Omega \partial_i
\Omega^\dagger = \Omega D_i \Omega^\dagger \hs , \hs D_i = \partial_i +
A_i 
\en
develops a magnetic monopole at the degeneracy points $x=\bar{x}$. 
\bi

\no
Defining the Abelian magnetic field by
\be
\label{E5}
{\cal B} _i = \epsilon_{ijk} \partial_j \lk A^{\Omega}_{i} \rk^3 
\label{eq:E5} 
\en
the magnetic flux through a closed surface $\Sigma$ is given by  
\bea 
\label{E6}
m & = & \int_\Sigma  d \vec{\Sigma} \vec{B}^3 = \il_\Sigma d \Sigma_i  
\epsilon_{ijk} \partial_j \lk A^{\Omega}_{k} \rk^3
\nonumber\\
& = & - 2 \il_\Sigma  d \Sigma_i  \epsilon_{ijk} \partial_j \tr \lk T^3
A^{\Omega}_{k} \rk \hk . 
\ena 
By Gau\ss' theorem this flux vanishes since $\partial \Sigma = 0$ 
unless the gauge potential is
singular somewhere on the surface $\Sigma$.
If the initial field configuration $A_i (x)$ was smooth, then the
first term 
in (\ref{E4}) will have no singularities. The magnetic monopole arises
from the second term in (\ref{E4}) ${\cal A } _i = \Omega \partial_i 
\Omega^\dagger $,  and only this term contributes to the
magnetic charge  (\ref{E6}), which hence can be written as  
\be
\label{E10}
m = - 2 \il_\Sigma d \Sigma_i \epsilon_{ijk} tr \lk T_3 \partial_j \lk
\Omega \partial_k \Omega^\dagger \rk \rk \hk . 
\en
For later convenience let us also introduce the normalized field $\hat{\Phi}
(x)$ defined by the unit vector in color space 
$\hat{\Phi} (\vec{x}) = \vec{\Phi}(\vec{x})/ \sqrt{\vec{\Phi}^2 (\vec{x})} $. 
Since $\hat{\Phi}(\vec{x}) \cdot \hat{\Phi} (\vec{x}) = 1$ 
this field defines a surface $S_2$ in color space. 
For given $\vec{ \Phi } (\vec{x})$ the field $\hat{\Phi }
(\vec{x})$ is well defined except at those points $x = \bar{x}$ where the
field  $\Phi (\vec{x})$ (\ref{eq:5}) vanishes. 
We have therefore to exclude the position 
$\bar x$ from the manifold on which  $\hat{\Phi}
(\vec{x})$ is defined, i.e. the field  $\hat{\Phi}
(\vec{x})$ is defined on the manifold $R^3 - \{\bar x\}$, which is
topologically equivalent to $S_2$.
Hence the chiral field  $\hat{\Phi} (\vec{x})$ 
defines a mapping from $S_2$ in coordinate space to the unit
sphere $S_2$ in color space, defined by  $\hat{\Phi}
(\vec{x}) \cdot  \hat{\Phi}(\vec{x}) = 1$. Since $\Pi_2 (S_2) = {\Zset}$ 
such mappings fall into homotopy
classes characterized by the winding number
\be
\label{E9}
n \left[ \hat{\Phi} \right] = \frac{1}{8 \pi} \il_{S_2}  
d^2u \;  \epsilon_{\alpha \beta } \hat{\Phi}
\cdot \lk \partial_\alpha \hat{\Phi} 
\times \partial_\beta \hat{\Phi} \rk
 \hk , 
\en 
where $u^{\alpha =1,2}$ are coordinates of the $S_2$ in coordinate space 
and $\partial _\alpha = \partial / \partial u^\alpha $. 
In the gauge
(\ref{E2}) this field has the representation 
\be
\label{E8}
\hat{{\bf \Phi }} (\vec{x}) \; = \; \hat{\Phi ^a}(\vec{x}) T^a \; = \; 
\Omega^\dagger (\vec{x}) T_3 \Omega (\vec{x}) \hk . 
\en
where $\Omega(\vec{x}) \in SU(2)/U(1) \simeq S^2$. With this
representation it is straightforward to show that the magnetic
flux (\ref{E10}) divided by the volume of the unit sphere
$S_2$, $4\pi$, coincides with the  winding number (\ref{E9}), 
i.e.\  $m / 4\pi = n[\hat{\Phi }] $. This implies that 
the magnetic charge of the monopole is quantized in integers. 
This result was already found in~\cite{kro87}. Note, however, 
that our definition of the magnetic charge differs from that 
in~\cite{kro87} by a factor of two. 

Let us emphasize that the quantization of the magnetic monopoles 
arising in the Abelian gauge of non-Abelian gauge theories already 
occurs at classical level on purely topological grounds. This is 
different from the 
Dirac quantization of magnetic monopoles in QED. In the latter case, 
the quantization of magnetic charge results in the presence of an integer 
electric charge from the uniqueness of the wave function . 

In order to calculate the magnetic charge (\ref{E10}), we have tacitly 
assumed that the gauge transformation $\Omega (\vec{x})$ is 
well defined in coordinate space except at the point $x=\bar{x}$. 
A glance at (\ref{eq:5}) shows that this is not the case, but 
$\Omega $ is ill-defined along a string attached to the monopole, 
which is just the pendant of the Dirac string in classical 
electro dynamics\footnote{The above evaluation of the magnetic flux 
of the monopole assumes that the point where the string pierces the 
surface is excluded. Inclusion of this point would yield zero net 
magnetic flux out of the sphere, since the flux of the string is 
oppositely equal to the total flux of the magnetic monopole, see 
also below.}. To be precise, the vector $\hat{\alpha }^a$ 
is singular for $\Phi ^3 + \Phi = 0$. 
Let us notice that this condition is in fact three constraints, which 
define the region in coordinate space where the gauge 
transformation (\ref{eq:4}) is not defined, i.e. 
\be 
\Phi ^1 (\vec{x}) =0 \; , \hbo 
\Phi ^2 (\vec{x}) =0 \; , \hbo 
\Phi ^3 (\vec{x}) \leq 0 \; . 
\label{eq:6} 
\en 
Whereas the first two equations generically select a line in space, 
the last condition of (\ref{eq:6}) chooses segments on this 
line. We therefore conclude that the regions where the gauge 
transformation (\ref{eq:4}) is ill-defined corresponds to strings 
which in most cases\footnote{Details will be presented in the next 
section.} will have finite length. At the end-points of these strings, 
the equality holds in the last constraint of (\ref{eq:6}) implying that 
Abelian monopoles occur at the endpoints of the string. 
By performing an Abelian gauge transformation $\Omega \to t\Omega,~ t \in 
H$, the string can be arbitrarily deformed. 

Let us illuminate the emergence of the string in more detail. 
For this purpose, we confine ourselves to the region of space 
close to the Abelian monopole. As discussed before, the gauge 
fixing field 
$\hat{\Phi }$ (\ref{E3}) then takes a hedgehog form. Without loss of 
generality, we may choose 
\be 
\hat{\Phi }^a \; = \; \hat{x}^a \; , \hbo 
\alpha ^a \; = \; r \delta ^{a3} + x^a \; , \hbo 
\hat{\alpha }^a \; = \; \frac{ r \delta ^{a3} + x^a }{ \sqrt{ 2r (r+x^3)} } 
\; , 
\label{eq:7} 
\en 
where $r^2=x^a x^a$. 
In this case, the gauge transformation (\ref{eq:4}) is ill-defined on 
the negative $x^3$-axis. 
In order to extract the ${\cal B}$-field configuration which is induced 
by the transformation (\ref{eq:4}), we regularize the induced gauge 
potential by 
\be 
{\cal A}^a_k (\vec{x}) 
\; = \; - 2 \tr \{ \Omega \partial _k \Omega ^{\dagger } 
\, T^a \} \; = \; \lim _{\epsilon \rightarrow 0} 
\frac{2}{ \alpha ^2 + \epsilon ^2 } \, \epsilon ^{abc} 
\alpha ^b \partial _k \alpha ^c \; . 
\label{eq:8} 
\en 
Introducing polar coordinates $(r, \theta , \varphi )$, the Abelian 
magnetic field (\ref{eq:E5}) is given for any value of the regulator 
$\epsilon $ by 
\be 
\vec{\cal B} ^3 \; = \; \frac{4 r^2 (1 + \cos \theta )^2 }{ 
[ 2 r^2 (1 + \cos \theta ) \, + \, \epsilon ^2 ]^2 } \, 
\vec{e}_r \; + \; \frac{ 4 \epsilon ^2 }{ 
[ 2 r^2 (1 + \cos \theta ) \, + \, \epsilon ^2 ]^2 } \, \vec{e}_3\; , 
\label{eq:9} 
\en 
where $\vec{e}_r$, $\vec{e}_3$ are the unit vectors in radial and 
$x^3$-direction, respectively. 
\begin{figure}[t]
\centerline{ 
\epsfxsize=9cm
\epsffile{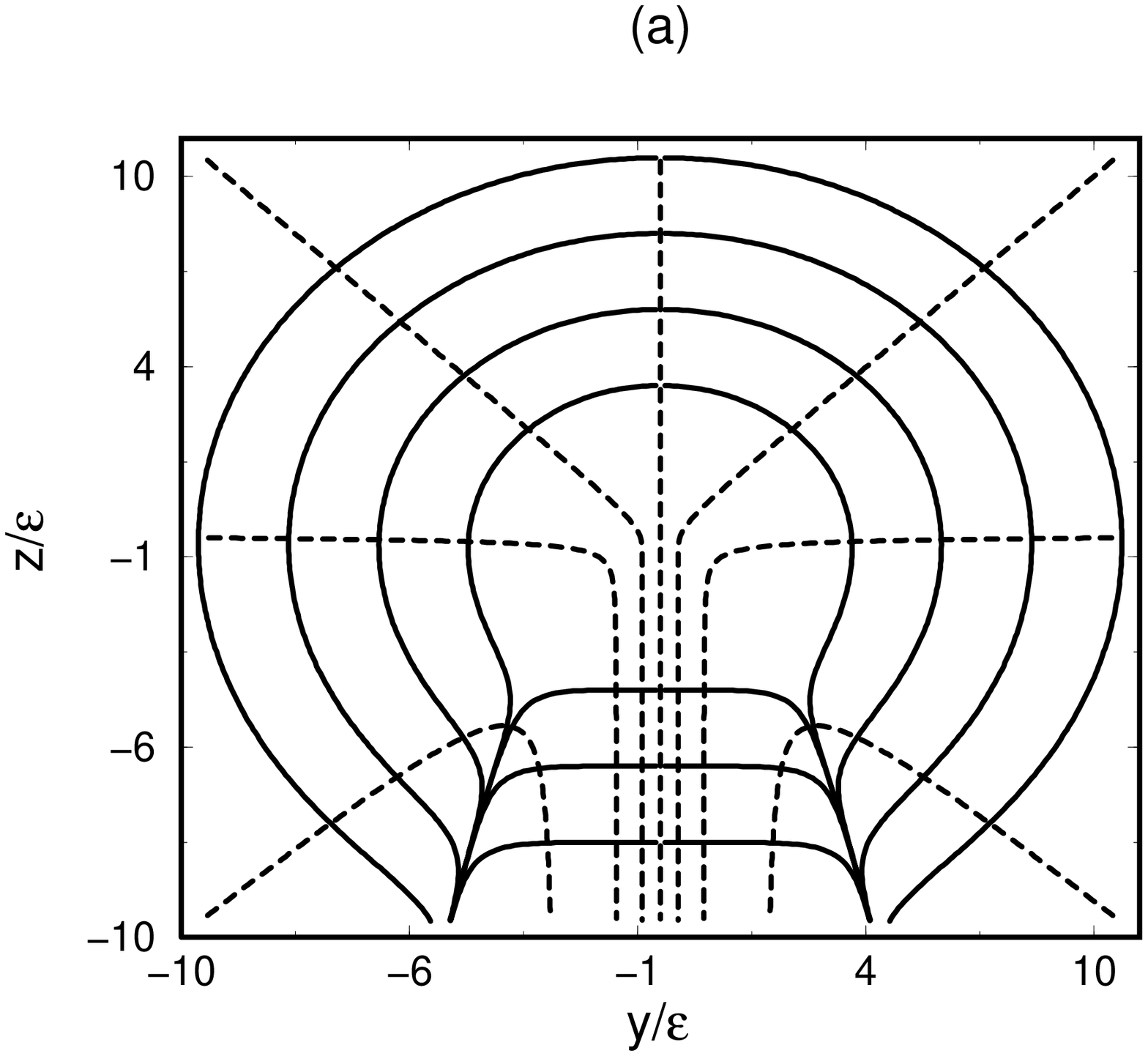} 
\epsfxsize=9cm
\epsffile{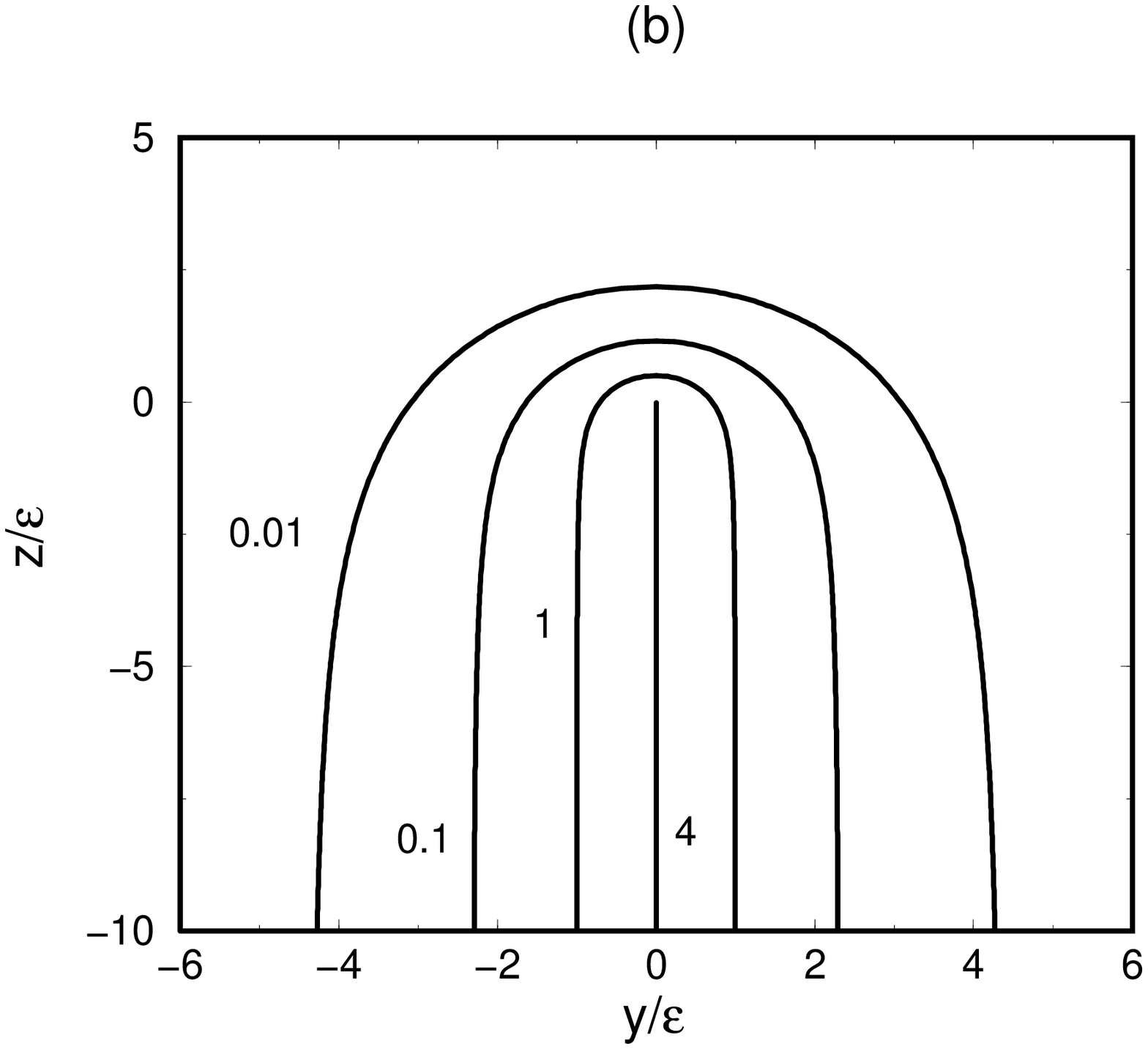} 
}
\vspace{.5cm} 
\caption{ The field lines (dashed) and the lines of constant strength 
   (solid) of the Abelian magnetic field $\vec{\cal B}^3 $ (a) 
   and the full color magnetic field $\vec{B}^3 $ (picture (b)) of the 
   regularized gauge field (\protect{\ref{eq:8}}). } 
\label{fig:1} 
\end{figure} 
Removing the regulator $\epsilon $ in (\ref{eq:9}), one easily verifies 
by calculating the magnetic flux through the plane $x^3 = const.$ that 
the second term in (\ref{eq:9}) yields a $\delta $-function for 
$x^3<0$, i.e. 
\be 
\vec{\cal B} ^3 \; = \; \frac{1}{r^2 } \, \vec{e}_r \; + \; 
4 \pi \, \delta ^{(2)}(\vec{x}) \, \theta (-x_3) \, \vec{e}_3 \; , 
\label{eq:10} 
\en 
where $\theta (x) $ is the step-function. The first term at the 
r.h.s.\ of (\ref{eq:10}) is precisely the Abelian monopole field. The 
second term represents the magnetic string lying on the negative 
$x^3$-axis. From eq.(\ref{eq:10}), it is obvious that the magnetic flux 
of the monopole flowing out of the sphere equals the flux of the magnetic 
string flowing into the sphere.

\section{ Monopoles versus strings }
\bi

\no
As discussed above, in the gauge $\vec{\Phi} = \Phi \tau_3$ the 
Abelian magnetic field ${\cal B}_i$ (\ref{E5}), arising from 
the induced gauge potential $\cA_i = \Omega \partial_i \Omega ^\dagger$, 
develops 
a magnetic monopole at those points at which $\Phi = 0$. 
In fact, we have seen that the employed gauge transformation is such that 
the quantity $\cA_i$ gives rise to a magnetic monopole field with a string 
attached. Contrary to classical electrodynamics, in the 
present case the string is not an artifact of a coordinate singularity but
physically meaningful. Let us explain this fact in more detail. 
\bi

\no
In classical electrodynamics the primary quantity is the magnetic field 
$\vec{\cal B}$ of the monopole satisfying the equation
\be
\label{70}
\vec{\nabla} \vec{\cal B} = g \delta^{(3)} (x) \hk , 
\en
while the gauge potential $\vec{\cal A}$ itself has no physical meaning. 

The (Dirac) string arises when one tries to represent the magnetic 
field of the monopole by a single overall defined gauge potential 
$\vec{\cal A}$ 
\be
\label{71}
\vec{\cal B} = \vec{\nabla} \times \vec{\cal A} \hk . 
\en
In this case, since $\vec{\nabla} \cdot \vec{\nabla}  \times \vec{\cal A} =
0$ for any regular function, the gauge potential $\vec{\cal A}$ 
has to be singular in order to produce a magnetic monopole field (\ref{70}), 
and the Dirac string arises. As we discussed in the previous section, 
the magnetic flux flowing through the Dirac
string into the monopole is the same as the flux of the monopole field,
and the monopole field together with the Dirac string gives rise to 
continuous magnetic flux lines, that is the total
magnetic field of the monopole and the Dirac string is source, free $
\vec{\nabla} \vec{\cal B} = 0$. In this case, one has to exclude the 
magnetic field of the Dirac string in order to be left with the net 
magnetic monopole field satisfying (\ref{70}).
In this sense, the Dirac string is not a physical object in classical 
electrodynamics\footnote{In fact, the emergence of the Dirac string 
can be avoided by using the Wu-Yang construction, which employs different 
gauge potentials in the upper $(\hat\Phi^3>0)$ and lower 
$(\hat\Phi^3<0)$ hemispheres. In this case the gauge potential
is discontinuous across the $\hat\Phi^3 = 0$ plane, and it is this
discontinuity in the gauge potential which then provides the net
magnetic flux of the monopole field. Note also that the Wu-Yang construction 
arises in a different gauge, namely $\hat{ \Phi } = \vec{e}_3 
\hbox{sign} ( \Phi ^3 )$. }. 

On the other hand, at quantum level the gauge potential itself becomes 
physically meaningful. It is this quantity which defines the quantum 
theory in both the operator and the functional integral approach. 
Furthermore, there are phenomena in topologically non-simply connected 
spaces, such as the Bohm-Aharonov effect, which cannot be exclusively 
described in terms of field strengths, but require resort to the gauge 
potential. Hence, in the quantum theory the string of the monopole field 
cannot be discarded. 
\bi

\no
Let us now consider the non-Abelian gauge theory and assume that 
we initially work in a gauge (e.g.~$A_0=0$) in which all fields 
configurations are smooth. 
The magnetic monopoles here arise when we bring (originally smooth) 
gauge configurations into a
particular (Abelian) gauge. By gauge invariance of the theory, 
the new gauge potential $A^{\Omega}_{\mu}$ is equivalent to the smooth
starting gauge potential $A_\mu (x)$ and there is no reason why one
should exclude parts of the gauge rotated potential  $A^{\Omega}_{\mu}$. In
particular, if the inhomogeneous part $\cA_i = \Omega \partial_i 
\Omega^\dagger$ develops a magnetic monopole, there is no reason to 
exclude the corresponding string. 

The string contributes to the Abelian magnetic flux (\ref{E6}) 
and cancels the contribution from the monopole (point) singularity 
as is easily checked by inserting (\ref{eq:10}) into 
(\ref{E5},\ref{E6}). 

Let us emphasize that the magnetic monopoles arise only in the 
Abelian magnetic field, which is induced by ${\cal A}_i= 
\Omega \partial _i \Omega ^{\dagger }$. At quantum level, there 
is a priori no reason to consider the Abelian magnetic field only. 
This is because the weight of a gauge potential in the partition 
function is determined by its full (non-Abelian) field strength. 
However, the induced gauge potential $\cA_i = \Omega \partial_i 
\Omega^\dagger$ is a pure gauge in that region of coordinate space 
where $\Omega (\vec{x})$ is well defined. Consequently, 
its field strength 
\bea 
B^a_k &:=& \frac{1}{2} \epsilon _{klm} F^a_{lm} 
\; = \; \frac{1}{2} \epsilon _{klm} \left[ 
\partial _l A^a_m \, - \, \partial _m A^a_l \, + \, 
\epsilon ^{abc} A^b_l A^c_m \right] 
\label{eq:11} \\ 
&=& {\cal B}^a_k - \frac{1}{2} \epsilon _{klm} \epsilon ^{abc} 
A^b_l A^c_m \; . 
\nonumber 
\ena 
vanishes except at those points where $\Omega (\vec{x})$ is singular.
Thus we expect a non-zero color-magnetic field $B_i$ only along the string 
singularity, and in the remaining part of the space the commutator term
in the field strength $\left[\cA_i, \cA_j \right]$ has to compensate the
Abelian magnetic field $\vec\nabla \times \vec \cA$. 
This can be explicitly demonstrated by using the 
regularized gauge potential $\cA_i$ (\ref{eq:8}) of the
monopole field and calculating the full (non-Abelian) magnetic field
strength (\ref{eq:11}), see fig. 1.  The monopole field has in fact 
disappeared, leaving only an extended string-like magnetic field. In the 
limit where the regulator is removed, one obtains a string of magnetic flux. 
Thus, instead of a gas of Abelian monopoles, we are left with an ensemble of 
strings. 

Obviously, the magnetic strings transform homogeneously under regular 
gauge transformations. In this respect, the flux strings are more 
convenient variables than the monopole potentials. Lattice calculations 
have revealed that in the maximal Abelian gauge magnetic monopoles 
are the dominant infra red degrees of freedom at least for 
some observables, e.g. the tension of the chromo-electric string 
between quarks in the fundamental representation~\cite{pol95}. 
Since any magnetic monopole is connected to a string, we can 
alternatively consider these strings as the infra red dominant degrees 
of freedom. In fact, on a lattice magnetic monopoles are identified 
by measuring the magnetic flux of the corresponding string~\cite{gra80}. 

\bi

\no 
To summarize this section, 
we have shown that a string of magnetic field strength 
is attached to the monopoles in the Abelian magnetic fields which 
arise in Abelian gauges. In addition, one finds that the monopole 
field cancels in the full color-magnetic field, whereas the color-magnetic 
string survives. As long as we assume that in the quantum theory 
the gauge potential is the primary quantity 
(and not the field strength, as is argued in classical 
electro dynamics), the string must be regarded as an effective 
degree of freedom. Lattice theories which were designed to focus 
on the role of the Abelian monopoles by adopting proper gauges, 
have revealed that these monopoles play an important role for the 
confinement mechanism. Knowing about the intimate relation between the 
strings and the monopoles in particular Abelian gauges, it is 
tempting to assume that the magnetic strings are in fact the dominant 
infra red degrees of freedom. Thus, instead of a gas of monopoles, 
we are left with an ensemble of magnetic strings. To investigate 
the properties of such a string-dominated Yang-Mills vacuum as well as 
to study its impact for confinement is an interesting task 
for future work. 

\section{ String statistics } 

From $Z_N$ vortex condensation theory~\cite{tho79}, we expect 
a non-trivial contribution of a string to the Wilson loop whenever 
the string pierces the loop area. It is therefore of particular 
interest to study the length distribution of the strings in the 
Yang-Mills vacuum. Since we do not know the field configurations 
which dominate the ground state, we must resort to a model description. 
At this stage of the investigation, we assume for simplicity that the 
Yang-Mills vacuum is given by a dilute instanton gas. 
In what follows, we will study the strings 
which arise by casting the instanton medium into an Abelian gauge. 

In order to be specific, we choose the following gauge, 
\be 
{\bf \Phi } (\vec{x}) \; = \; 
\int _{-\infty }^{\infty } dt \; F^2[A] \; A_0(t,\vec{x}) \; = \; 
\Phi (\vec{x}) \; T^3 \; , 
\label{eq:12} 
\en 
where $F^2[A]$ is the squared field strength tensor evaluated with the 
gauge field configuration $A_\mu (x)$ under consideration. 
Literally, eq.\ (\ref{eq:12}) is the time average of the zeroth component 
of the gauge potential, where $F^2$ acts as a measure for the 
averaging procedure. It is obvious that a time independent gauge 
transformation $\Omega (\vec{x})$ suffices to cast an arbitrary 
(time dependent) configuration into the gauge (\ref{eq:12}). 
Hence, the results 
of the previous sections apply also in the present case. 
Note that different powers of the field strength tensor in (\ref{eq:12}) 
are possible as well. Different choices correspond to different gauges. 

The model configuration whose string content we wish to study here 
is a gas of $N$ (anti-) instantons, i.e. 
\be 
A^a_\mu (x) \; = \; \sum _{k=1}^{N/2} O_{ab}^{(k)} 
\frac{ \eta ^b _{\mu \nu } (x_\nu - x_\nu ^{(k)} ) }{ 
(x_\nu - x_\nu ^{(k)} )^2 + \rho ^2 } \; + \; 
\sum _{k=1}^{N/2} \bar{O}_{ab}^{(k)} 
\frac{ \bar{ \eta }^b _{\mu \nu } (x_\nu - \bar{x}_\nu ^{(k)} ) }{ 
(x_\nu - \bar{x}_\nu ^{(k)} )^2 + \rho ^2 } \; , 
\label{eq:13} 
\en 
where $O_{ab}^{(k)}$ ($\bar{O}_{ab}^{(k)}$) is an orthogonal matrix 
which defines the orientation of the (anti-) instanton, and 
$x_\mu ^{(k)}$ ($\bar{x}_\mu ^{(k)}$) is the position of the $k$th 
(anti-) instanton. We have assumed that the instantons possess 
an unique radius $\rho $, which is not an unrealistic 
assumption~\cite{shu84}. 
Note that the configuration (\ref{eq:13}) 
is an approximate solution of the Yang-Mills equation of motion if 
the instanton distances are large compared with $\rho $. 
Here, we do not want to construct the lowest action solution which 
might serve as candidate for the (semi-classical) vacuum of 
Yang-Mills theory, but we would like to discuss the properties 
of the magnetic strings arising from a medium which is somewhat close 
to the (semi-classical) ground state. 

\begin{figure}[t]
\parbox{8cm}{ 
\hspace{1cm} 
\centerline{ 
\epsfxsize=8cm
\epsffile{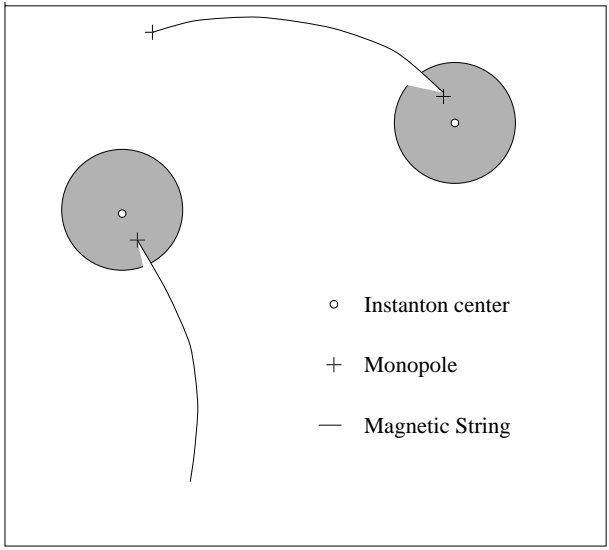} 
}
} \hspace{1cm}
\parbox{5cm}{ 
\caption{ Schematic figure showing the magnetic strings produced 
   by two instantons and the correponding monopoles. } 
} 
\label{fig:2a} 
\end{figure} 
For technical simplification, we approximate the total squared  
field strength in (\ref{eq:12}) by the superposition of the squared 
field strength of the single instantons, i.e. 
\be 
F^2[A] \; \approx \; \sum _{k=1}^{N} \frac{ \rho ^4 }{ 
\left[ (x_\nu - x_\nu ^{(k)} )^2 + \rho ^2 \right] ^4 } \; . 
\label{eq:14} 
\en 
Again, this approximation is justified, if the instanton gas is dilute. 
Alternatively, we can interpret the use of 
(\ref{eq:14}) in (\ref{eq:12}) as a change of the gauge choice. 

\begin{figure}[t]
\centerline{ 
\epsfxsize=8.5cm
\epsffile{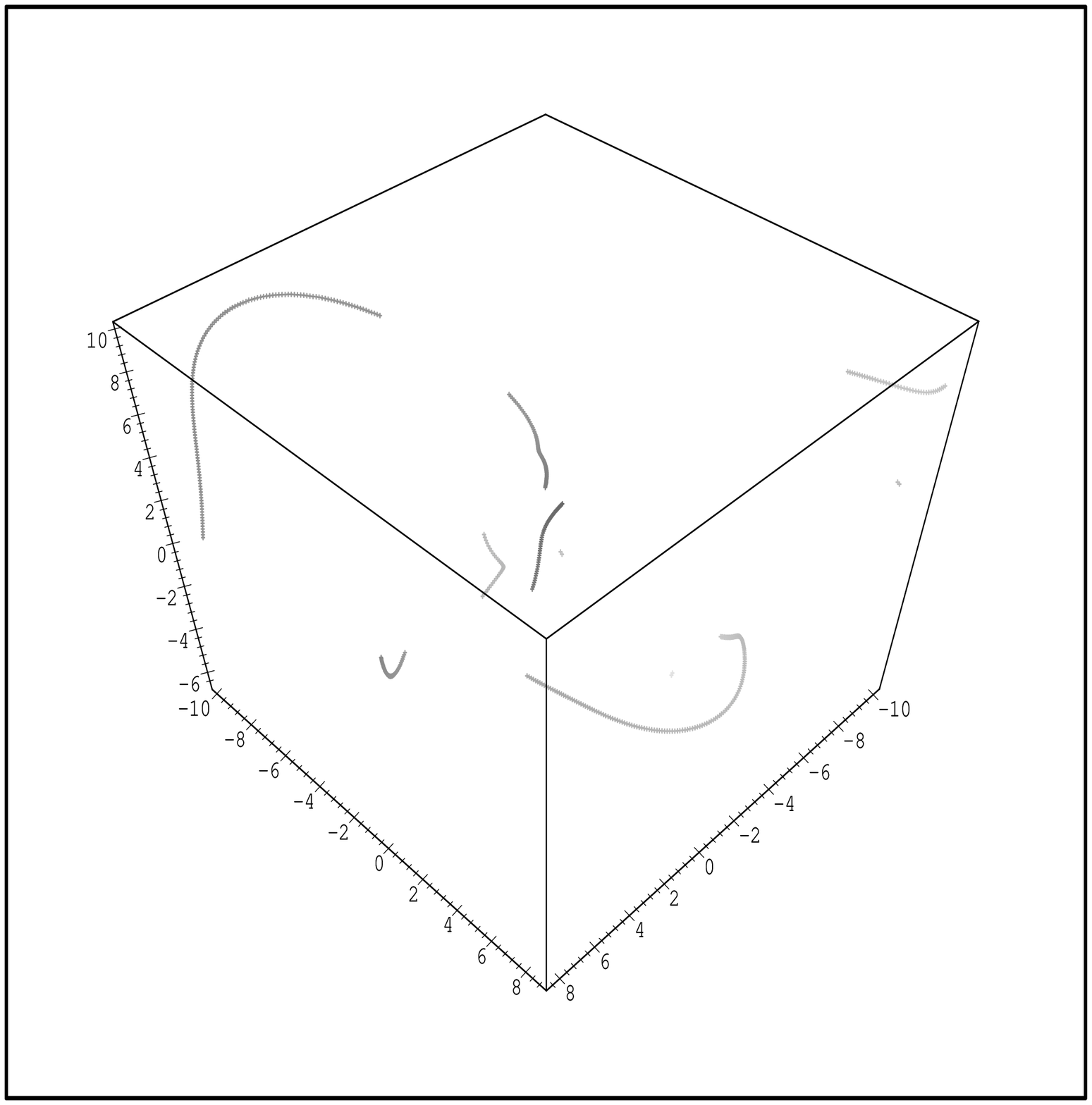} 
\epsfxsize=8.5cm
\epsffile{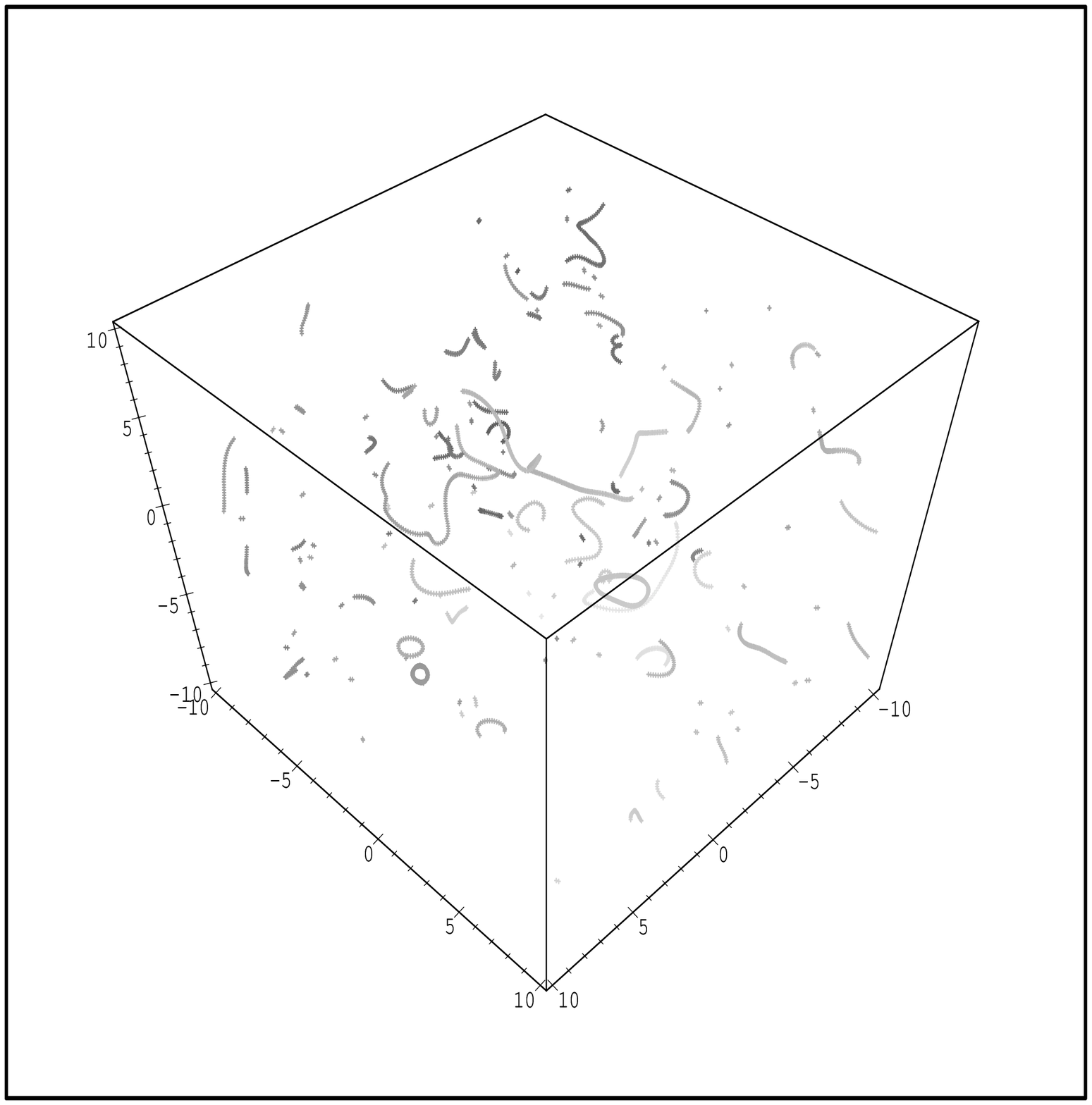} 
}
\vspace{.5cm} 
\caption{ The color-magnetic strings in coordinate space 
   for $N=10$ (left) and $N=300$ (right) instantons inside 
   a box of side length $20 \, \rho$. }
\label{fig:2} 
\end{figure} 
For a single instanton, the magnetic monopole occurs at the instanton 
center, and the string extends from the center to infinity. 
Let us also briefly discuss the case of two instantons. 
For two widely separated instantons, the monopole position is given by 
the solution of the equation 
\be 
\vec{\Phi }_1(\vec{x}) \; + \; \vec{\Phi }_2(\vec{x}) \; = \; 0 \; , 
\; \label{eq:a1} 
\en 
where $\vec{\Phi }_i(\vec{x})$ is the functional $\vec{\Phi }$ 
(\ref{eq:12}) 
evaluated with the i-th instanton configuration. Since the contribution 
of $\vec{\Phi }_2(\vec{x})$, evaluated at the center of the first instanton, 
to the above equation is small, we expect a monopole occurring close 
to each instanton center. In addition, one finds a further monopole 
in between the two instantons. This additional 
monopole occurs, where both functions $\vec{\Phi }_i(\vec{x})$ in 
(\ref{eq:a1}) are small and cancel each other. 
Figure \ref{fig:2a} qualitatively shows the behavior of the magnetic 
string, if two instantons are present. The generic picture for many 
instantons is that a string of finite length is attached to a 
point close to each instanton center. 

In order to be precise, 
we have calculated the positions of the strings arising from a 
multi-instanton medium, 
when the quantity (\ref{eq:12}) satisfies the conditions (\ref{eq:6}). 
Figure \ref{fig:2} shows the color-magnetic strings in coordinate space. 
The calculation was done for $N=10$ and $N=300$ instantons, 
and the positions and orientations of the instantons are chosen randomly. 
One observes that if the instanton density increases the length of the 
strings decreases. 

\begin{figure}[t]
\centerline{ 
\epsfxsize=8.5cm
\epsffile{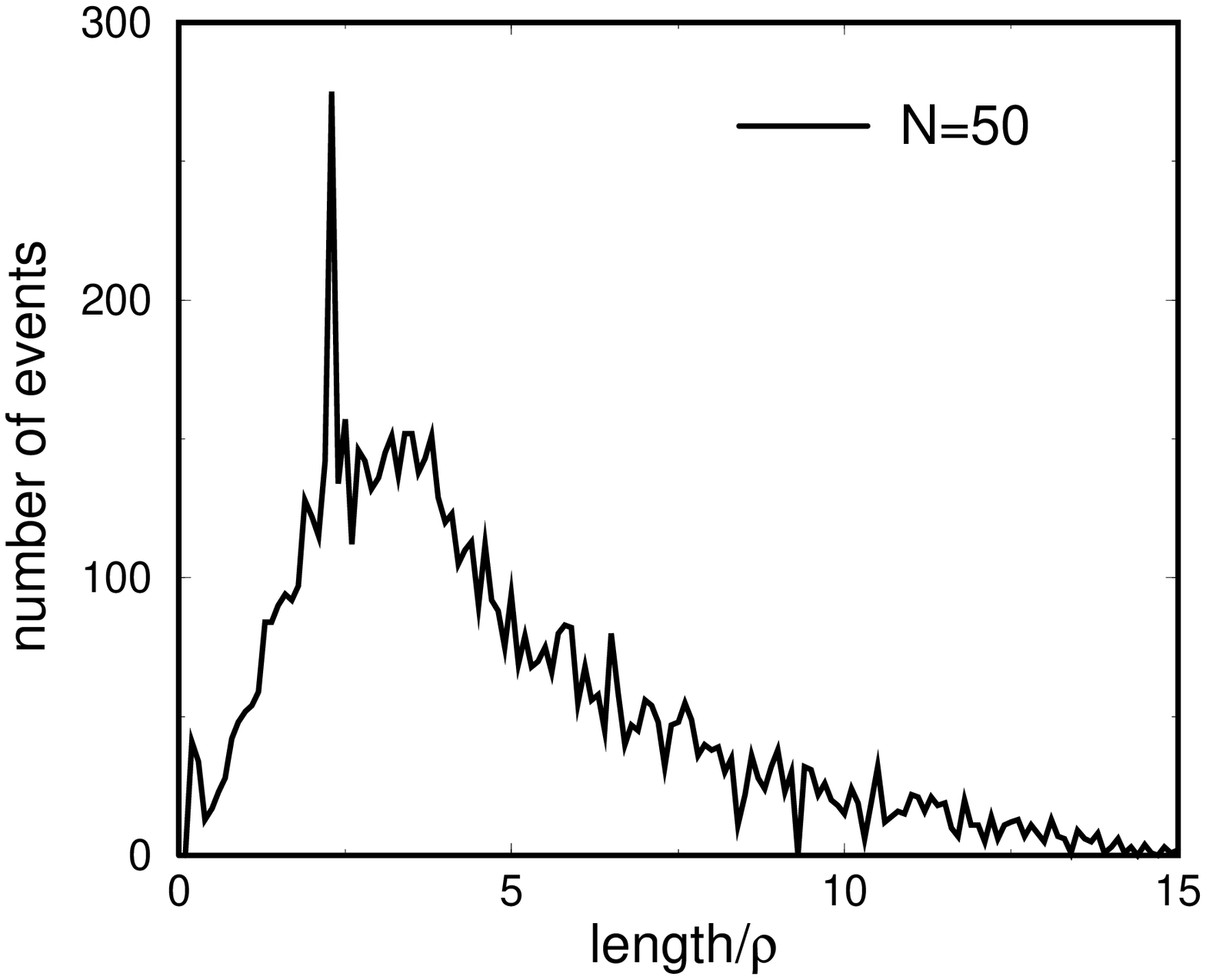} 
\epsfxsize=8.5cm
\epsffile{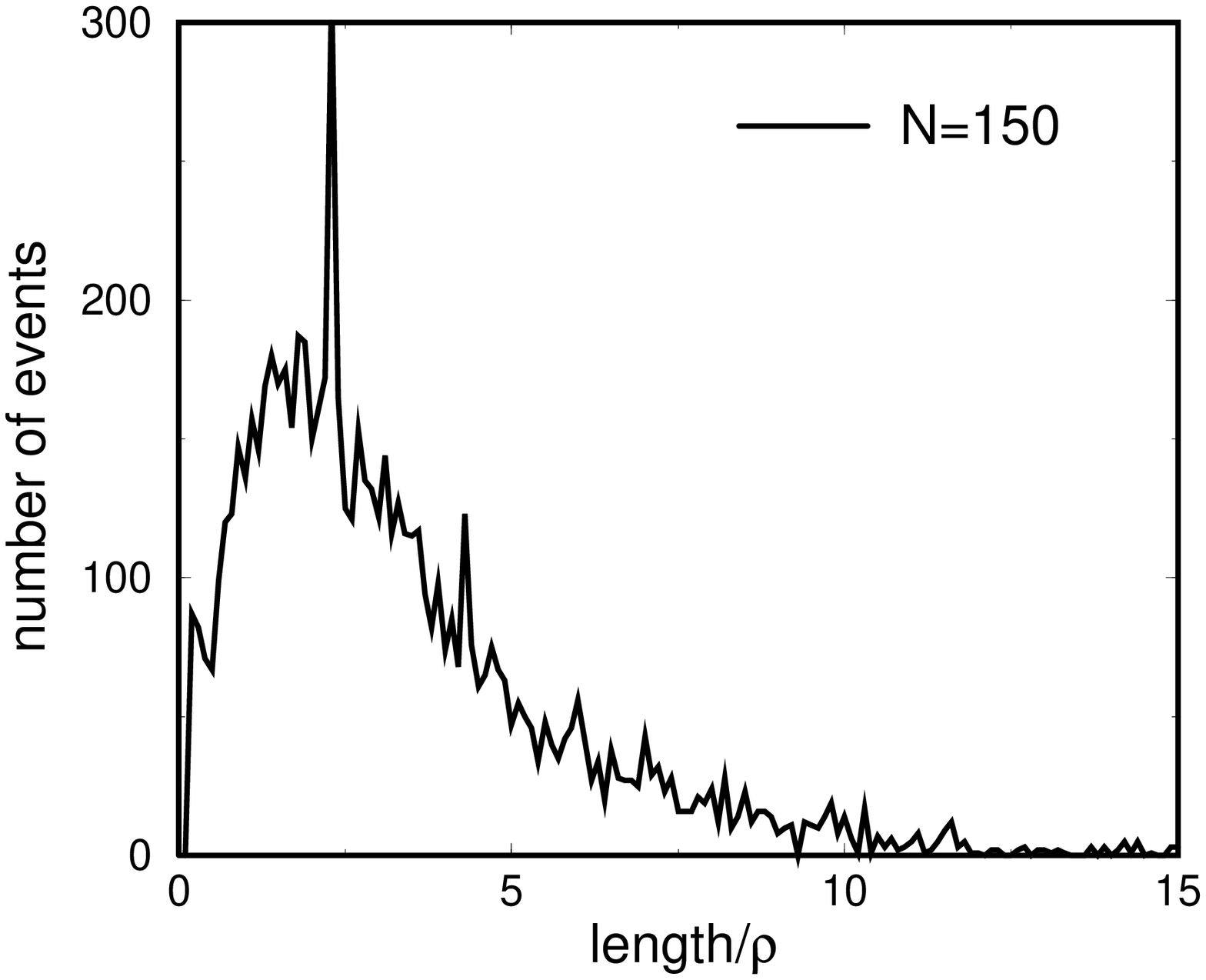} 
}
\vspace{.5cm} 
\caption{ The distribution of the string length for $N=50$ (left) 
   and $N=150$ (right) instantons. 
}
\label{fig:3} 
\end{figure} 
By investigating several thousands of samples of strings inside the 
box, we extracted the string length distribution. The result is 
shown in figure \ref{fig:3}. One observes a peak in the distribution 
at $length \approx 2.3 \, \rho $ which is independent of the 
instanton density, whereas the bulk of the distribution strongly 
depends on number of instantons inside the box. The peak structure 
could arise from string configurations the length of which is determined 
by the properties of a single instanton rather than by the 
average instanton distance, e.g. by closed strings looping around 
the instanton center. 

\begin{figure}[t]
\parbox{8cm}{ 
\hspace{1cm} 
\centerline{ 
\epsfxsize=8cm
\epsffile{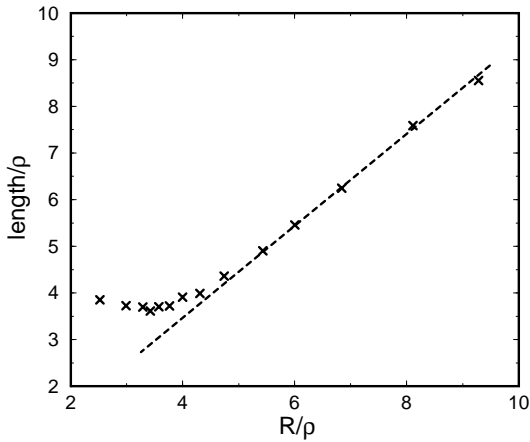} 
}
} \hspace{1cm}
\parbox{5cm}{ 
\caption{ The average length of the color-magnetic strings as 
   function of the average inter-instanton distance $R$. } 
} 
\label{fig:4} 
\end{figure} 
Finally, we present the relation between the string length and 
the inter-instanton distance (figure \ref{fig:4}). At small densities 
(large $R$), the string length scales with the average distance $R$ 
between the instantons. If this average distance is smaller than 
a critical distance $R_c \approx 4 \, \rho $, the string length 
almost stays  constant. A possible explanation is the following: at small 
instanton densities, the string length associated with a particular 
instanton is limited by the neighboring instanton. If the density 
increases, the strings start to percolate between the instantons, and 
therefore their lengths depend on the instanton radius rather than on $R$. 
Although specific properties of string configurations might change, if 
the dilute gas approximation is abandoned, we do not expect 
significant changes in the gross features of the string distribution. 

Let us comment on the physical implications of the 
string distribution. 
If a magnetic string pierces the minimal surface spanned by a 
Wilson loop, we expect a non-vanishing contribution to Wilson's integral. 
The average length of the strings is therefore of crucial importance, 
since it determines the average number of strings piercing the surface. 
In a forthcoming publication, we will 
study the contribution of the strings to the Wilson loop in a 
quantitative manner.

\section{ Conclusions } 

We have studied the emergence of magnetic monopoles and strings in 
the Abelian projection (see e.g.~\cite{pol95}), which had been brought  
into question by recent papers~\cite{dam95,deb96,deb94}. 
We have argued that magnetic strings are the relevant infra red 
dominant degrees of freedom rather than monopoles. 
This conjecture is based on two observations: firstly, Abelian 
monopoles are an artifact in the sense that they always appear 
whenever a residual U(1) gauge degree of freedom remains unfixed. 
Secondly, magnetic strings rather than monopoles appear in the non-Abelian 
field strength. The monopole in the Abelian magnetic field, which stems 
from the neutral gauge field $A^3_\mu $, is canceled in the full 
non-Abelian field strength by contributions from the charged 
fields $A^{ch}_\mu $. This implies that the charged gauge potentials 
carry as much topological information as the Abelian field $A^3_\mu$. 

The above conjecture received support from lattice 
calculations~\cite{deb96}. It was observed that the string tension 
can be reproduced by assuming a ''center dominance''. Within this 
approximation, the residual U(1) gauge-symmetry on the lattice 
is further reduced to a local $Z_2$ symmetry, and the ground 
state of such a lattice model is known to be dominated by vortices 
rather than by monopoles. 

In this paper, we did 
not further pursue the phenomenology of a string dominated ground 
state---this is left to a forthcoming publication---, but 
studied the properties of the strings which arise from a dilute 
instanton gas. Although this is not a very realistic model for the Yang-Mills 
vacuum~\cite{sh82}, it should suffice to obtain first insights into 
string properties at least at finite temperatures. 
We found that the string length scales with the 
average instanton distance at small instanton densities, whereas 
the length depends only on the instanton radius at higher densities. 
An interesting issue for future investigations would be the 
identification of the strings on the lattice, and the study of their 
impact on confinement. 

\bigskip 
\no
{\bf Acknowledgments:} 

\no
We thank M.~Engelhardt for useful comments on the manuscript.

\end{document}